# Lung Cancer Detection from CT Scan Images based on Genetic-Independent Recurrent Deep Learning

Ehsan Sadeghi Pour
*Department of Electrical and Computer Engineering*
*Islamic Azad University, Kashan Branch*
Kashan, Iran
Ehsan61.ai@gmail.com

Mahdi Esmaeili
*Department of Electrical and Computer Engineering*
*Islamic Azad University, Kashan Branch*
Kashan, Iran
M.esmaeili@iaukashan.ac.ir

*Abstract*— **Lung cancer is one of the prevalence diseases in the world which cause many deaths. Detecting early stages of lung cancer is so necessary. So, modeling and simulating some intelligent medical systems is an essential which can help specialist to accurately determine and diagnose the disease. So this paper contributes a new lung cancer detection model in CT images which use machine learning methods. There are three steps in this model: noise reduction (pre-processing), segmentation (middle-processing) and optimize segmentation for detect exact are of nodules. This article use some filters for noise reduction and then use Independent Recurrent Neural Networks (IndRNN) as deep learning methods for segmentation which optimize and tune by Genetic Algorithm. The results represented that the proposed method can detect exact area of nodules in CT images.**

## I. Introduction

Lung cancer is one of the most important issues in the medical world that can be identified using artificial intelligence techniques. Lung cancer is the most common cancer in the world and is currently considered an epidemic in the global scale [1]. In the near future, it is expected that the incidence of these violent events will increase. The disease is also recognized as the cause of death among various types of cancers, so that 86% of cases died within 5 years after diagnosis, causing 29% of all deaths from cancer. Although different methods for its detection and control are currently being used [2]. The main factor is smoking in terms of risk factors for this disease. Carcinogenic substances and the underlying cause of the tumor in cigarettes increase the risk of developing primary lung cancer, so that 85% of patients with any type of lung cancer with any type of histology had previous history of smoking (previous or current) [3]. The relative risk of developing this cancer by smoking is about 13 times, and in case of prolonged passive smoking it also increases 1.5 times. Additionally, exposure to certain industrial compounds such as arsenic, asbestos and chromium, people with chronic pulmonary disease, and previous history of tuberculosis, with secondary scarring, are associated with an increased risk of developing lung carcinoma [3].

New calculations from the International Agency for Research on Cancer represented that world-wide deaths from cancer in 2007 were more than 1.8 million, and it will likely to reach 10 million deaths per year in 2030 which the highest increases would be in developing countries (70% of deaths from cancer). On the other hand, attention to the major contribution of lung cancer to overall cancer deaths (1.3 million deaths per year) and its dependence on smoking, points out that primary lung carcinoma is a major health issue with more adverts are generally bad for the world and hence the recognition of the status and trend of lung cancer and its changes in recent years can pave the way for planning to prevent and even predict the necessary future health care for lung cancer [3]. The major problem associated with cancer is a lack of timely diagnosis or in general, a weakness in the diagnosis of the disease, which is also due to the lack of proper pattern selection by the physician or the lack of proper use of standard patterns. Therefore, implementing a method that can help each person to properly diagnose the disease can be a major step in preventing and controlling the disease, especially in its early stages.

Based on the experimental results of WebMD [2], which is one of the lung cancer organizations, about ¼ of people have no symptoms at the time their lung cancer has been diagnosed, and this diagnosis is based on the use of x-rays. The data are classified into three output classes, namely patient class, health, and suspicious. With the aid of data classification, data classify into classes with the same characteristics. In general, classifiers are divided into supervised and unsupervised categories. In the supervisor classification, data have tagged and its attributes are specified. Unsupervised classifications do not include tagged data. In this approach based on methods, unsupervised learning used for classification. Lung cancer is one of the most common malignancies in the world, which is an uncontrolled cell growth in lung tissue [1] and is the cause of the highest cancer death rate in men and the second leading cause of cancer death in women (after breast cancer) [2].

Lack of patient complaints and clinical signs in the early stages of the disease as well as low sensitivity of laboratory methods are the most important causes of death in lung cancer. Pulmonary nodules are a common symptom of lung cancer. Computed tomography (CT) technology is a good tool for diagnosing the lungs, which is very difficult for a radiologist to examine a few millimeters in diameter due to the large volume of CT images; Therefore, Computer Aided Diagnosis (CAD) methods are being developed every day as an important and practical tool to help radiologists in this matter [1-3]. The five-year survival rate for patients with lung cancer after surgery is only 14%, while early detection when the glands are small increases the patients' chances of survival by 70% to 80% [2]. Pulmonary nodules can appear separately or attached to the lung wall. The more regular the pulmonary nodules, the more benign they are, and the more irregular and vague the border, the more malignant they are. At present, areas are identified as nodules or non-nodules, and a specialist can definitely determine whether the area is nodular or non-nodular. However, the diagnosis of benign and malignant pulmonary nodules by a physician is based on probability and he can not 100% judge the benignity or malignancy of pulmonary nodules and give a definite opinion. Malignant nodules can attach to a number of arteries in the lung tissue, and the more blood the nodule feeds on the arteries, the more malignant the nodule. Malignant nodules in the lung indicate lung cancer. The size and probability of benign and malignant can be estimated from the size of the masses, so that the probability of malignancy of pulmonary nodules with a size less than 4 mm 0%, between 7-4 mm 1%, between 8-20 mm 15% and larger It is 75% from 20 mm. Masses larger than 1 cm are easily visible to the naked eye [1-3]. Computer-aided detection methods can be divided into two categories: methods based on Intensity Based brightness and methods based on Model Based model. In the model-based method, we already have information about the shape of the nodule pattern, and therefore methods such as pattern matching, models based on anatomy, and morphology, the shape of the nodules are in this category. The pattern of pulmonary nodules is obtained from experimental information or educational data (8). In the diagnostic process, two types of errors can occur: Negative errors are related to items that are actually nodules but are not categorized as nodules; Positive error due to incorrect reporting of non-nodular cases (vessels, wounds, etc.) is called nodule. Prerequisite for a diagnostic assist system is high sensitivity and low positive error rate. Among the important advantages of the proposed method, all structures with a brightness level of 1 and close to the edge of the image are eliminated. This increases the speed and accuracy of processing by limiting the search space, and its sensitivity in detecting cancerous nodules and masses will increase compared to the sensitivity of the methods used. Boundary delimitation has always been an important issue in medical imaging [3]. Therefore, this article proposes a new method of lung cancer detection by optimized segmentation.

The rest of his paper organized as below. In literature review, some recent methods survey for lung cancer diagnosis via images and clinical datasets. Proposed method represents a new approach for lung cancer detection based on Independent Recurrent Neural Networks (IRNN) and Genetic Algorithm. Then a simulation has done in MATLAB environment with CT images which contributed by as standard dataset. At the end, a conclusion provided to overall review of his article.

## II. LITERATURE REVIEW

There are many uses of machine learning algorithms in the field of health to predict the risk of disease, diagnosis and anomaly, and the likelihood of responding to treatment. For example, data mining has been used to predict the survival rate of patients. For example, in [4] presented a prediction model for the survival rate of transplanted or transplanted kidney tissue, but in the prediction model, machine learning techniques were used and showed that the machine learning algorithms were compared to advanced statistical methods such as Cox-regression-based Nomo-grams provide better results.

In [5] data mining algorithm used to analyze the inconsistency of a clinical guide to the treatment of lung cancer, and finally obtained four rules using an Inductive Decision Tree. Finding patterns of non-compliance with the clinical guidelines by using these rules is a good alternative to manual methods. The obtained rules can be used in the knowledge base of decision-making systems based on a clinical guide to alert in cases of non-compliance with the clinical guidelines. Another study in [6] proposed a system to predict response to treatment using actual lung cancer data. In this proposed system consisting of the following four steps: 1) data collection 2) feature selection 3) design of the classification model 4) computational parameters that ultimately used the combination of filter and wrapper algorithms and the embedded method. Then, the proposed method was compared with the linear model of Hazards to evaluate the algorithm, and it was found that the combination model had better results.

In [7], the predication radiation pneumonitis in locally advanced stage II–III non-small cell lung cancer using machine learning proposed. The combined predictive performance of radiation pneumonitis predictors such as maximum esophagus dose, lung V20, mean lung dose, pack-year, lung V5 and lung V10 improves the performance of individual predictors up to a 24.6% improvement rate using random forest. Another method which presented in [8], used image processing for lung cancer by optimal deep learning as classification. CT Scan image dataset used as inputs. Modified Gravitational Search Algorithm is applied to train the Optimal Deep Neural Network. The proposed classifier provides the sensitivity of 95.26%, specificity of 96.2% and accuracy of 96.2%.

Lung cancer patient survival via supervised machine learning classification techniques predicted in [9]. The models performed well with low to moderate lung cancer patient survival times. This approach created a custom ensemble, enabling evaluation of each model's predictive power. Results of the models are consistent with a classical Cox proportional hazards model. The mechanistic understanding and the tools for effective prevention, early diagnosis, and therapy of lung cancer described in [10]. Also assessing indoor air pollution exposure and lung cancer risk in Xuan Wei, China studied in [11].

Deep learning methods and algorithms used recently for lung cancer diagnosis with image data and clinical datasets. For example in [12], Convolutional Neural Network (CNN) used for predicting melanoma and lung cancer from histopathology images. In another example whih proposed in [13], two deep learning techniques RedNet and ProNet used for lung cancer classification in 3D CT scan images. In [14] a new kinds of neural network name two-step deep learning method used for lymph node diagnosis and is metastases from histopathology images. Another deep learning technique used for predicting lung cancer risk in CT images [15]. This study can improve the early detection of lung cancer and surveyed its initial risks. Also in [16], a new deep learning approach proposed for lung cancer detection via clustering-based method entitles Improved Profuse Clustering Technique (IPCT) in CT images. Lung cancer diagnosis with CNN proposed in [17], too. In this article, a new classification method based on CNN provided from CT images to represent the stage of disease. Another study which proposed in [18], used a combination algorithm based on deep-reinforcement learning in an Internet of Things environment and platform for new kinds of lung cancer detection. This article provided a smart medicine and electronic health system in hospitals and clinics. Predicting lung cancer stages especially T1/T2 proposed in [19] in CT images with deep learning method CNN. Also in [20], P2.16-02 features in T1 stage predicted from pathology images with 3D deep CNN. Another features of lung cancer in histopathology images P1.09-32 used for prediction and classification with deep CNN [21]. In the other side, many researcher wrote an article for predicting MA20.10 from clinical datasets with deep CNN [22].

Another deep learning method used for lung cancer diagnosis and classification, too, such as U-Net [23], Convolutional Long Short Term Memory (C-LSTM) [24], Unsupervised Deep Fuzzy C-Means clustering Network (UDFCMN) [25], Deep Gene Coexpression Network [26], ResNet [27], Deep-Reinforcement Learning [28], 3D CNN [29], Fully Convolutional Network (FCN) [30], 3D Deep Residual Network [31], and so on. For performing nodule segmentation, in [32] proposed an algorithm. The algorithm used two-region growing methods for performing segmentation of nodule which are contrast-based region growing and fuzzy connectivity. The local adaptive segmentation algorithm is applied for identification of background and foreground regions within a defined window size. This method performed better for a separated nodule but, to segment the attached nodules, the algorithm failed. In [33] developed an algorithm that adopted geodesic influence zones in image of multi-threshold representation. This allows a fusion segmentation criterion that is based on both object nodule segmentations itself. In [33] used morphological and threshold techniques for elimination of background and other surrounding information from the provided region of interest. For classification of each pixel in the detected space, the researchers used support vector machines. Similarly, in [34] segmented the solitary pulmonary nodules in digital radiography (DR) images. This is achieved by incorporating a sequential filter for construction of new representations of the weight and matrices of probability. This work is limited to DR images which limits the usage for CT images. Additionally, in [35] worked on ground-glass nodules segmentation. The researcher used an asymmetric multi-phase deformable model. In this method, robustness was the constraint for addressing the requirements of segmentation for other types of nodules.

The notion of optimally clustering a set of feature vectors was designed by [36]. The process involves extraction of shape and intensity related features in a given feature data space from a predicted nodule. The size is calculated by measuring the segmented nodule volume via an ellipsoid approximation by making use of equivalent diameters of the segmented regions in a 2.5-D representation. Hence, there exists uncertainty in the final results. In [37] developed a scheme which is based on voxel intensity segmentation. It incorporates mean intensity based thresholding in the Geodesic Active contour model in level sets. This work was validated on limited set of scans, so the robustness of the proposed technique is dubious. In reference [38] the literature on breast cancer detection and classification based on ML algorithms was reviewed. In reference [39] A hybrid of the firefly algorithm and artificial intelligence (AI) was employed to detect breast cancer. In [40] proposed the work which cuckoo search has been hybridized with the clustering algorithm. This algorithm has been introduced to understand the protein cell machinery activity with the help of protein complexes. This idea helps in detecting the protein complexes, which is very difficult with the existing traditional algorithms. The authors have detected the protein complex cores in each and every dynamic sub- networks and this has been achieved after the construction of a dynamic protein networks. At the final phase, the attachments of protein have been grouped with CS to their own cores. In [41] have been worked on CS algorithm, on the satellite images with the aim of contrast enhancement. This work was specifically focusing on the enlargement of the virtual effects of the satellite images. Authors have considered the minimum procurement and also its natural conservation which contains the minimal image artefacts. They tried with a new enhancement algorithm which will helps in enhancement of satellite stills. This work carried in three different threads. During the initial pass, the adaption of chaotic initialization was done and this could result in the image's premature convergence. After this, during the second pass, fitness value has been adapted which significantly turned on strategy of adaptive Levy Flight, to seek the result of improvement in the CSA convergence rate. At the final pass, introduced a mutation randomization technique for the felicitation and effective balancing to be incorporated between exploration and so called exploitation threads, and hence ensured with the best feasible optimum solution. In references [42] and [43], the XCSR and XCSLA classification system for diagnosing some of the diseases are that it can also be used to diagnose breast cancer. In [44] for the effective fragmentation of the biometric image, the authors have worked on the active contour module. In this work, authors have presented a model to segment biomedical image, by considering intensity inhomogeneity, which imposed Local Mean and variance (LMV). The technique called as Local Mean and Variance (LMV) Active Contour. In this proposed model, the assumption of Gaussian distribution has made over the distribution of intensity that is represented in each region. Meanwhile, energy function was derived and introduced in various levels. By acquiring the energy minimization, the curve evolution formula was obtained. In [45] Authors have proposed a theory on automated 3-D lung segmentation technique that was formulated using active-contour model, on CT lung images. The proposed system was integrated the active contour model (ACM) along with the local image bias field formulation. Here, mean squared errors which are a local energy formula have been used to accommodate in homogeneous CT images and finally detecting and fragmenting tumor region. This is very effective with intensity inhomogeneity technique. Smoothening this process, CT images were subjected to a Multi-Scale Gaussian Distribution and this leads to the features determination of the image. Also in [46], Cuckoo Search Optimization (CSO) algorithm applied to optimize the initial segmentation of the lung portion, and then Active Contour (AC) technique has been applied to segment the nodules from the segmented lung image. In order to fine turn the post- processing after the nodule segmentation operation, Markov Random Field (MRF) technique has been applied. In [47], lungs cancer classification from CT images proposed which used an integrated design of contrast based classical features fusion and selection in three steps: 1) gamma correction, 2) multiple texture, point, and geometric features for feature extraction, and 3) zero values and negative features are replaced by an entropy-based approach. In another study [48] AI and image processing techniques were employed to detect breast cancer.

In study [49] evaluate a noise reduction for mammographic pictures and to identify salt and pepper, Gaussian, and Poisson so that precise mass detection operations can be estimated. In [50] a new model for detect and extract tumor area from MRI images proposed. At first, some pre-processing method applieddue to enhancing input image which selected from BraTS dataset; Then, segmentation of MRI images applied by morphological operators which represented image edges and edema as two main features besidetexture, light intensity and shape. Then cellular neural network training started to learn features from segmented part which could detect exact area of tumors. Lungs Data Science Bowl 2017 used as input datasets [51].

## III. PREPARE YOUR PAPER BEFORE STYLING

This study uses an independent deep recurrent neural network to segment CT images to detect pulmonary nodules. The deep neural network approach of this research is recursive, but in this structure called RNN, there is the problem of slope disappearance and slope explosion (slope is the same as gradient). Therefore, to solve this problem, another structure called independent recursive neural network (IndRNN) is considered in this research, which solves the problems of disappearance and gradient descent in a traditional fully connected RNN. Each neuron in a layer receives only its past state as background information (instead of being fully connected to the other nerve cells in that layer) and so the neurons are independent of each other's history. To prevent the gradient from fading and exploding in order to preserve long-term or short-term memory, the slope cross-diffusion rate can be adjusted. The information of the reciprocal neurons is examined in the following layers. IndRNN can be strongly trained with unsaturated nonlinear functions such as ReLU. Deep grids can be trained using jump connections. The overall structure of the independent deep recurrent neural network can be described as equation (1).

$$h_t = \sigma(Wx_t + u \odot h_{t-1} + b) \quad (1)$$

Where the recursive weight $u$ is a vector and $\odot$ represents the Hadamard product. Each neuron is independent of the other layers in one layer, and communication between neurons can be achieved by placing two or more layers of independent deep recurrent neural network. For $\odot$ 'th neurons, the latent state $h_{n,t}$ can be obtained as equation (2).

$$h_{n,t} = \sigma(w_n x_t + u_n h_{n,t-1} + b_n) \quad (2)$$

Where $w_n$ and $u_n$ are the ninth row of input weight and repeat weight, respectively. Each neuron receives information from its input and hidden state only at the previous time stage. That is each neuron in an independent deep recurrent neural network independently deals with a kind of spatio-temporal pattern. Typically, the recursive deep neural network acts as a multilayer receptor over time as parameters are shared. The proposed independent recursive deep neural network differs from the conventional recursive deep neural network, providing a new perspective on recursive neural networks as an independent analysis of spatial patterns (i.e. via $W$ over time or e.g. via $u$. Correlations between different nerve cells can be used by placing two or more layers, in which case each neuron in the next layer processes the output of all the neurons in the previous layer. This steps can solve segmentation via training and testing of CT images. For better segmentation, Genetic Algorithm applied for tuning IndRNN.

At the post processing stage, Genetic Algorithm has been applied to fine tune the segmented nodule by considering some agents such as initial population of chromosomes, elite genes, crossover, mutation and random selection in an iteration by considering a fitness function as equation (3).

$$E(u) = \sum_{i \in V} D(u_i) + \sum_{(i,j) \in E} V(u_i, u_j) \quad (3)$$

## IV. SIMULATION AND RESULTS

This article use Data Science Bowl 2017 dataset which described in [43]. Input dataset show in Fig. 1.

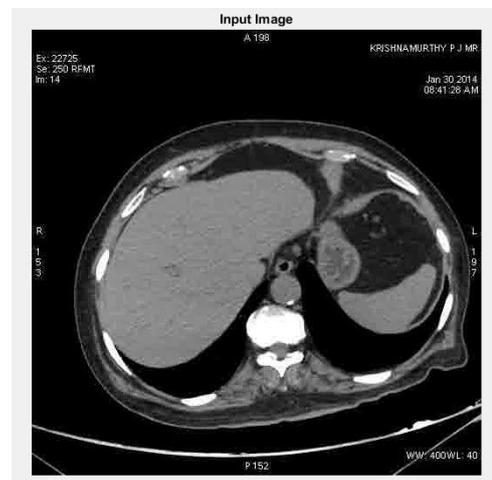

Fig. 1. Input image

Next, it is necessary to first pre-process in order to remove the initial noise of the image with the median filter, the results of which can be seen in Fig. 2.

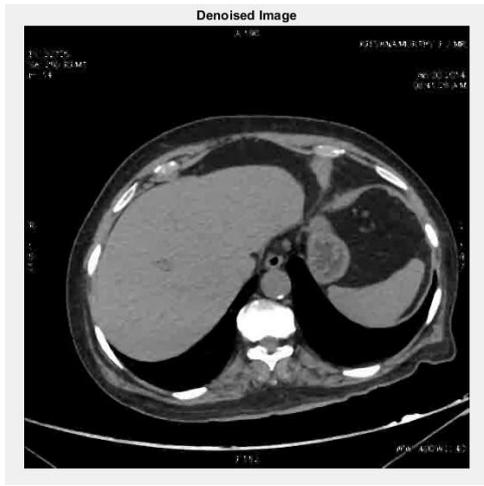

Fig. 2. Noise removal and reduction with median filter

Next, in the preprocessing section, light intensity conversions are performed, which results in Fig. 3 and 4, in which a type of filter is applied to the image in red so that more details can be seen.

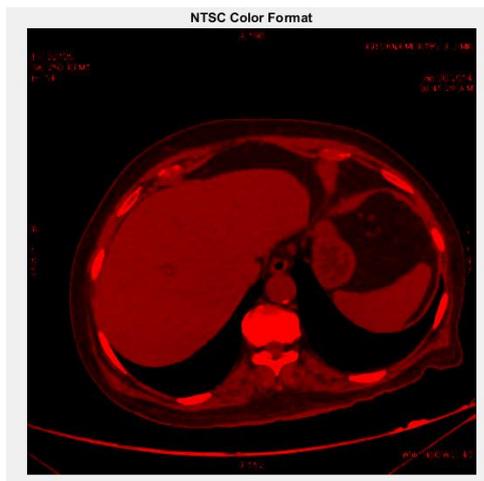

Fig. 3. Light intensity conversion

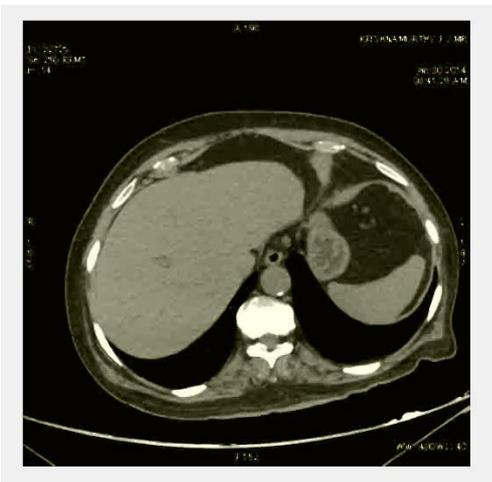

Fig. 4. Light intensity conversion

Initial segmentation steps for detecting lung nodules by IndRNN represented in Fig. 5.

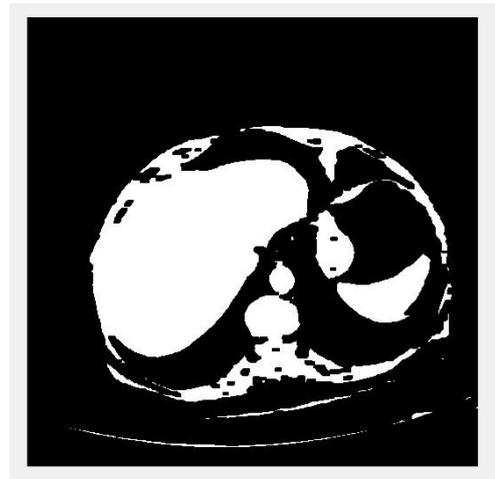

Fig. 5. IndRNN initial segmentation

Genetic algorithm operators are then applied to local IndRNN-based fractions, resulting in Fig. 6.

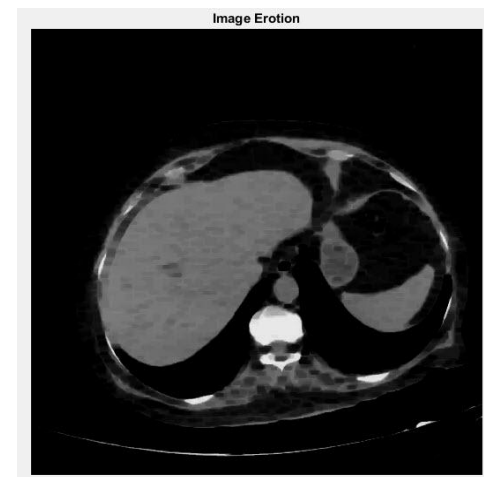

Fig. 6. Genetic algorithm operator applied for better segmentation (chromosome as initial population)

Based on the structure of the genetic algorithm embedded in this study to improve IndRNN, which works as a growth area based on local features such as texture, edge and light intensity in this study, a specific area of pulmonary nodule is extracted, which is illustrated in Fig. 7. This step done by considering 100 chromosomes and 5 elite genes with 0.2 crossover rate and 0.02 mutations per 100 iteration rounds with a random selection function.

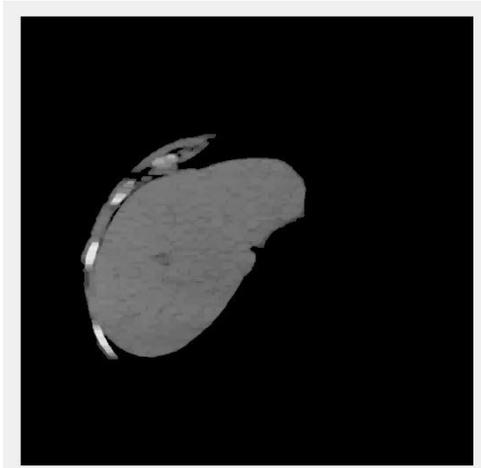

Fig. 7. Genetic algorithm applied for tuning IndRNN

A windowing operation is performed in this section, which is in the form of 8×8 and applied in the genetic algorithm. This windowing is applied to the results of Fig. 7 where the area of the pulmonary nodule is marked. The result is also shown in Fig. 8.

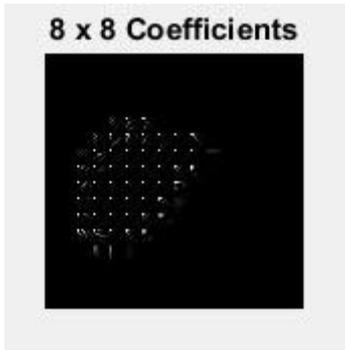

Fig. 8. Windowing operation

Finally, with the combined approach of IndRNN and improving its structure with a windowed genetic algorithm, pulmonary nodule segmentation is performed with the aim of lung cancer detection from CT scan images, the final result of which is shown in Fig. 9 and 10.

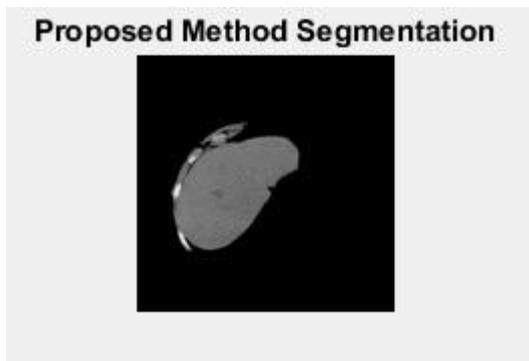

Fig. 9. Final results of lung nodules

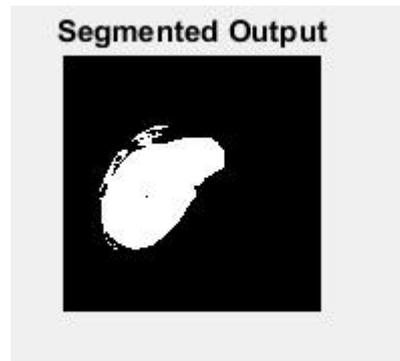

Fig. 10. Final results of lung nodules in binary mode

The evaluation results considered in this study are the criteria of accuracy, sensitivity and specificity that are obtained when implementing the proposed approach. The results can be presented in a case-by-case comparison with other methods, which are in terms of accuracy, sensitivity and specificity and are compared in Table 1. It should be noted that the comparison has scientific value when different methods such as reference articles [42] and [43] have used the same data set and this research has used the same data set. In other words, comparisons should be made under the same conditions, but the methods should be different.

TABLE I.  EVALUATION CRITERIA COMPARISON

| Refs. | Specificity | Sensitivity | Accuracy |
|---|---|---|---|
| Spoorthi Rakesh, et al., 2021 [46] | 79.81 % | 68.01 % | 89.21 % |
| M. Attique Khan, et al., 2020 [47] | 98.2 % | 89.32 % | 95.2 % |
| Proposed Method | 98.5 % | 91.25 % | 98.97 % |

Also Fig. 11 represented coefficient of similarity and spatial overlapping in comparison to [42]. Results represented that proposed method coefficient of similarity is 0.789 and spatial overlapping is 0.456, but proposed method in [42] is 0.009 and 0.358, respectively.

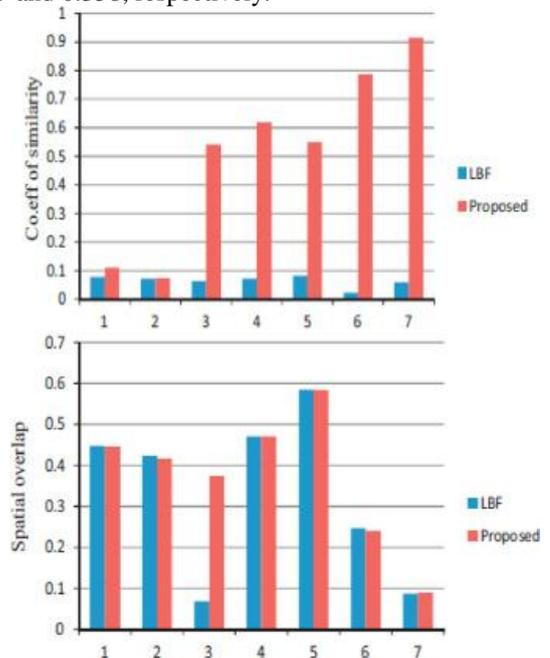

Fig. 11. Coefficient of similarity and spatial overlapping

## V. CONCLUSION

This paper proposed a new method of lung nodule segmentation for lung cancer detection from CT images by Data Science Bowl 2017 datasets. This algorithm consists of three steps: 1) noise reduction and removal by median filter, 2) segmentation with IndRNN as deep leaning method, and 3) tuning and optimizing IndRNN with Genetic Algorithm. Comparative analysis has been reported that proposed model is close to optimal and so better than [46] and [47] in terms of evaluation criteria such as accuracy, sensitivity, specificity, coefficient of similarity and spatial overlapping.